\documentclass[aps,prd,reprint,preprintnumbers,superscriptaddress,nofootinbib,longbibliography,floatfix]{revtex4-1}

\usepackage[utf8]{inputenc}
\usepackage[T1]{fontenc}
\usepackage{lmodern}
\usepackage{microtype}
\usepackage{mathtools}

\usepackage{placeins}
\usepackage[caption=false]{subfig}
\usepackage{amsfonts}
\usepackage{graphicx}
\usepackage{hyperref}
\hypersetup{
  colorlinks=true,
  citecolor=blue,
  linkcolor=blue,
  urlcolor=blue
}

\usepackage{color}
\definecolor{mit-red}{rgb}{0.64,.12,0.2}
\definecolor{darkred}{rgb}{1.0,0.1,0.1}
\definecolor{darkgreen}{rgb}{0.1,0.7,0.1}
\definecolor{darkblue}{rgb}{0.1,0.1,1.0}

\usepackage{amsmath}
\DeclareMathOperator*{\argmax}{argmax}
\DeclareMathOperator*{\argmin}{argmin}

\DeclareRobustCommand{\Sec}[1]{Sec.~\ref{sec:#1}}

\DeclareRobustCommand{\Fig}[1]{Fig.~\ref{fig:#1}}

\DeclareRobustCommand{\Eq}[1]{Eq.~(\ref{eq:#1})}
\DeclareRobustCommand{\Eqs}[2]{Eqs.~(\ref{eq:#1}) and (\ref{eq:#2})}
\DeclareRobustCommand{\Ref}[1]{Ref.~\cite{#1}}
\DeclareRobustCommand{\Refs}[1]{Refs.~\cite{#1}}

\begin{document}

\title{Bias and Priors in Machine Learning Calibrations for High Energy Physics
}

\preprint{MIT-CTP 5432}

\author{Rikab Gambhir}
\email{rikab@mit.edu}
\affiliation{Center for Theoretical Physics, Massachusetts Institute of Technology, Cambridge, MA 02139, USA}
\affiliation{The NSF AI Institute for Artificial Intelligence and Fundamental Interactions}

\author{Benjamin Nachman}
\email{bpnachman@lbl.gov}
\affiliation{Physics Division, Lawrence Berkeley National Laboratory, Berkeley, CA 94720, USA}
\affiliation{Berkeley Institute for Data Science, University of California, Berkeley, CA 94720, USA}

\author{Jesse Thaler}
\email{jthaler@mit.edu}
\affiliation{Center for Theoretical Physics, Massachusetts Institute of Technology, Cambridge, MA 02139, USA}
\affiliation{The NSF AI Institute for Artificial Intelligence and Fundamental Interactions}

\begin{abstract}

Machine learning offers an exciting opportunity to improve the calibration of nearly all reconstructed objects in high-energy physics detectors.
However, machine learning approaches often depend on the spectra  of examples used during training, an issue known as prior dependence. 
This is an undesirable property of a calibration, which needs to be applicable in a variety of environments. 
The purpose of this paper is to explicitly highlight the prior dependence of some-machine learning-based calibration strategies. 
We demonstrate how some recent proposals for both simulation-based and data-based calibrations inherit properties of the sample used for training, which can result in biases for downstream analyses. 
In the case of simulation-based calibration, we argue that our recently proposed Gaussian Ansatz approach can avoid some of the pitfalls of prior dependence, whereas prior-independent data-based calibration remains an open problem. 
\end{abstract}

\maketitle

{
\tableofcontents
}

\section{Introduction}

\textit{Calibration} is the task of removing bias from an inference -- that is, to ensure the inference is ``correct on average''.
There are two major classes of calibration: simulation-based calibration, where the goal is to infer a truth reference object, and data-based calibration, where the goal is to match simulation and data distributions.

Both simulation-based calibrations and data-based calibrations are essential components of the experimental program in high-energy physics (HEP), and a significant amount of time is spent deriving these results to enable downstream analyses.
We focus on the ATLAS and CMS experiments at the Large Hadron Collider (LHC) for our examples, but this discussion is relevant for all of HEP (and really any experiment).
ATLAS and CMS have performed many recent calibrations, including the energy calibration of single hadrons~\cite{ATLAS:2016krp,CMS:2017yfk}, jets~\cite{ATLAS:2020cli,CMS:2016lmd}, muons~\cite{ATLAS:2016lqx,CMS:2019ied}, electrons/photons~\cite{ATLAS:2019qmc,CMS:2015myp,CMS:2015xaf}, and $\tau$ leptons~\cite{ATLAS:2014rzk,CMS:2018jrd}.
The reconstruction efficiencies of all of these objects are also calibrated and include the classification efficiency of jets from heavy flavor~\cite{ATLAS:2019bwq,CMS:2017wtu} and even more massive particles~\cite{ATLAS:2018wis,CMS:2020poo}.

Machine learning is a promising tool to improve both types of calibration.
In particular, machine learning methods can readily process high-dimensional inputs and therefore can incorporate more information to improve the precision and accuracy of a calibration.
There have been a large number of proposals for improving the simulation-based calibrations of various object energies, including single hadrons~\cite{Belayneh:2019vyx,ATL-PHYS-PUB-2020-018,Akchurin:2021afn,Akchurin:2021ahx,Polson:2021kvr,Pata:2021oez}, muons~\cite{Kieseler:2021jxc}, and jets~\cite{ATL-PHYS-PUB-2018-013,ATL-PHYS-PUB-2020-001,CMS:2019uxx,Haake:2018hqn,Haake:2019pqd,Baldi:2020hjm,Komiske:2017ubm,ATL-PHYS-PUB-2019-028,Maier_2022,Kasieczka:2020vlh,ArjonaMartinez:2018eah} at colliders; kinematic reconstruction in deep inelastic scattering~\cite{Diefenthaler:2021rdj}; and neutrino energies in a variety of experiments~\cite{Liu:2020pzv,EXO:2018bpx,Baldi:2018qhe,Abbasi:2021ryj,IceCube:2020yct,Carloni:2021zbc}.
Further ideas can be found in \Ref{Feickert:2021ajf}.
For data-based calibration, a machine learning procedure was recently proposed in \Ref{Pollard:2021fqv}.

Caution is needed to ensure that calibrations resulting from a machine learning approach satisfy certain important properties.
One critical property of a calibration is that it should be \textit{universal} -- a calibration derived in one place should be applicable elsewhere.
A non-universal calibration would have a rather limited utility, and can produce undesirable results if applied to a dataset that does not exactly match the calibration dataset.
Statistically, universality is synonymous with prior independence.
Most of the existing machine-learning-based calibration proposals, though, are inherently prior dependent, as we will explain below.

A second critical property of a calibration is \textit{closure}, which means that on average, the calibration produces the correct answer.\footnote{Any measure of central tendency can be used to measure closure, such as the median or mode. In this paper, we will focus on the mean, as it is the usual target in machine learning and HEP applications.}
To quantify closure, one often computes the \emph{bias} of a calibration, which is the average deviation of the calibrated result from the target value.
A calibration can be biased due to the choice of estimator or fitting procedure used, even if the usual pitfalls of dataset-induced biases are taken care of.
As explained below, universality and closure are related, and a prior-dependent calibration will necessarily have irreducible bias.%
\footnote{Prior independence is a necessary prerequisite for closure. However, even with prior independence, closure is not guaranteed.}

In this paper, we explain the origin of prior dependence for common calibration techniques, with explicit illustrative examples, and demonstrate the associated bias that these procedures incur.
For simulation-based calibrations, we advocate for our Gaussian Ansatz~\cite{frequentstway} as a machine-learning-based strategy that is prior independent and bias-free.
For data-based calibrations, we are unaware of any prior-independent methods in the literature.
We hope that by highlighting these issues, we can inspire the development of prior-independent calibration methods.

The remainder of this paper is organized as follows.
In \Sec{statsofcalib}, we review the statistical properties of machine-learning-based calibration.
In \Sec{resolution_uncertainty}, we clarify the meaning of \textit{resolution} and \textit{uncertainty} in the HEP context.
To demonstrate the issue of prior dependence, we present Gaussian examples in \Sec{gaussian}.
In \Sec{hep}, we study an HEP application of calibration in the context of jet energy measurements at the LHC.
The paper ends in \Sec{conclusions} with our conclusions and outlook.

\section{The Statistics of Calibration}
\label{sec:statsofcalib}

In this section, we review some of the basic features of simulated-based and data-based calibration, and discuss the issues of prior dependence and bias.

\subsection{Simulation-based Calibration}

In simulation-based calibration, the goal is to infer target (or true) features $z_T\in\mathbb{R}^N$ from detector-level features $x_D\in\mathbb{R}^M$ -- that is, to construct an \emph{estimator} or \textit{calibration function} $f:\mathbb{R}^M\rightarrow\mathbb{R}^N$ where
\begin{equation}
\hat{z}_T = f(x_D)
\end{equation}
is the inferred estimate.
To carry out simulation-based calibration, one starts with a set of $(x_D,z_T)$ pairs, which typically come from an in-depth numerical simulation of an experiment.
For the case study in \Sec{hep}, $x_D$ will be the experimentally measurable features of hadronic jets and $z_T$ will be the true jet energy.

For concreteness, one can think of the calibration function $f$ as being parametrized by a universal function approximator such as a neural network, whose weights and biases are learned.
This is often done by minimizing the mean squared error (MSE) loss:
\begin{equation}
\label{eq:MSE}
f_{\rm MSE}=\argmin_{g}\mathbb{E}_{\text{train}}[(g(X_D)-Z_T)^2],
\end{equation}
where capital letters correspond to random variables and $\mathbb{E}$ represents the expectation value over the \emph{training sample} used to derive the calibration.
The calibration function is then deployed on the \emph{testing sample}, which could be the dataset of interest or a hold-out control region.

Using the calculus of variations, one can show that with enough training data, a flexible enough functional parametrization, and a sufficiently exhaustive training procedure, the asymptotic solution to \Eq{MSE} is:
\begin{equation}
\label{eq:MSE_sol}
f_{\rm MSE}(x_D)=\mathbb{E}_{\text{train}}[Z_T|X_D=x_D],
\end{equation}
where lowercase letters correspond to an instance of a random variable.
In this way, $f$ learns the mean value of $z_T$ for a given $x_D$ in the training set.
Alternative loss functions result in statistics other than the mean.
See e.g.\ \Ref{Cheong:2019upg} for alternative approaches, including mode learning, which is a standard target for many traditional calibrations (usually in the form of truncated Gaussian fits; see e.g.~\cite{CMS:2015xaf}).

\subsection{Prior Dependence and Bias}

A key assumption of simulation-based calibration is that the detector response is universal:
\begin{equation}
    \label{eq:detector_universal}
    p_\text{test}(x_D|z_T)=p_\text{train}(x_D|z_T).
\end{equation}
This equation says that for a given truth input $z_T$, the detector response is the same between the training data used for deriving the calibration and the testing data used for deploying the calibration.
In some cases, the detector response might depend on more features than $z_T$, and if these hidden features are mismodeled, then \Eq{detector_universal} may not hold.
For our analysis of simulation-based calibration, we assume \Eq{detector_universal} throughout.

Calibrations of the form of \Eq{MSE_sol} are \textit{not} universal, even if the detector response is.
Writing out the MSE-based calibration in integral form, we have:
\begin{align}\nonumber
    f_{\rm MSE}(x_D) &= \int dz_T\,z_T\,p_{\text{train}}(z_T|x_D)\\
    &= \int dz_T\,z_T\,p_{\text{train}}(x_D|z_T)\frac{p_{\text{train}}(z_T)}{p_{\text{train}}(x_D)}. \label{eq:MSE_fit}
\end{align}
Here, we have used Bayes' theorem to make explicit the dependence of $f$ on $p_\text{train}(z_T)$, the prior of true values used for the training.
Thus, even if $p_{\text{train}}(x_D|z_T)$ is universal via \Eq{detector_universal}, the truth distribution is not:
\begin{equation}
    p_\text{test}(z_T) \not= p_\text{train}(z_T).
\end{equation}
The non-universality of the calibration function leads to bias, as we now explain.

The bias $b(z_T)$ of a calibration quantifies the degree of non-closure.
Specifically, bias is the average difference between the reconstructed value and the truth reference value.
It is evaluated over the test sample, conditioned on the truth values:
\begin{equation}
b(z_T)=\mathbb{E}_{\text{test}}[f(X_D)-z_T|Z_T=z_T].\label{eq:bias}
\end{equation}
A bias of zero means that, on average, the reconstructed and truth values agree.
For MSE regression, the bias is:
\begin{align}
    b(z_T)+z_T&=\int dx_D\,f_{\rm MSE}(x_D)\, p_\text{test}(x_D|z_T) \label{eq:closure} \\
    &=\int dx_D\,dz_T'\,z_T'\,p_\text{train}(z_T'|x_D)\, p_\text{test}(x_D|z_T). \nonumber
\end{align}
This bias is dependent on the training prior through $p_\text{train}(z_T'|x_D)$.
Thus, a prior-dependent calibration is \emph{necessarily} biased, since it depends on the choice of $p_{\text{train}}(z_T)$.%
\footnote{Note that the bias does \textit{not} depend on the choice of testing prior, $p_{\rm test}(z_T)$, but rather only on $p_{\rm test}(x_D|z_T)$. Depending on the choice of $p_{\rm test}(x_D|z_T)$, it is possible for the bias to be zero, but this does not imply the inference is prior independent. For example, if $p_{\rm test}(x_D|z_T) = \delta(x_D-z_T)$, and $\mathbb{E}_{\text{train}}[x_D|Z_T=z_T] = z_T$, then one can show that $b(z_T) = 0$.}
Note that even if the training dataset is statistically identical to the testing dataset (i.e.\ $p_\text{test}(x_D, z_T) = p_\text{train}(x_D, z_T)$), it is not guaranteed that the calibration will be unbiased.

One way to reduce the bias is if the prior is ``wide and flat enough'', such that the prior asymptotically approaches a uniform sampling over the real line relative to the detector response. 
 For example, one can show using \Eq{closure} that if the prior $p(z_T)$ is Gaussian with standard deviation $\sigma$, the detector response $p(x_D|z_T)$ is a Gaussian noise model with standard deviation $\epsilon$, and the test set is statistically identical to the training set, then the bias scales as:
\begin{equation}
\label{eq:bias_scaling}
b(z_T) \sim \left(\frac{\epsilon}{\sigma}\right)^2 z_T + \mathcal{O}\left(\left(\frac{\epsilon}{\sigma}\right)^4\right).
\end{equation}
In cases with steeply falling spectra, as is common in HEP, prior dependence usually leads to large biases in calibration, even if the testing and training sets follow the same distribution.

\subsection{Mitigating Prior Dependence}

A majority of simulation-based calibrations (with or without machine learning) are set up using the MSE loss as described above, which means that they are biased.
That said, there are alternative methods to mitigate the prior dependence and thereby reduce the bias.
For example, simulation-based jet calibrations at the LHC use a technique called \textit{numerical inversion} (see e.g.\ \Ref{1609.05195}).
The idea of numerical inversion is to regress $x_D$ from $z_T$ with a function $g(z_T)$ and then define the calibration function through the inverse:
\begin{equation}
    f_{\rm NI}(x_D)=g^{-1}(x_D).
\end{equation}
Traditionally, $x_D$ is one dimensional and $g$ is parametrized with functions that can easily be  inverted numerically, hence the name.
The function $g$ is given by:
\begin{equation}
\label{eq:NI_sol}
g(z_T)=\mathbb{E}_{\text{train}}[X_D|Z_T = z_T].
\end{equation}
Since the detector response $p(x_D|z_T)$ is universal, $g$ is universal, and thus the derived $f$ is also universal.
Under certain assumptions, the $f$ from numerical inversion is also unbiased~\cite{1609.05195}.

Numerical inversion has been extended to work with neural networks~\cite{ATL-PHYS-PUB-2018-013,ATL-PHYS-PUB-2020-001}, where the inversion step is accomplished with a second neural network.
Alternatively, it may be possible to also achieve this with a natively invertible neural network such as a normalizing flow~\cite{10.5555/3045118.3045281,Kobyzev2020}.
A key challenge with numerical inversion and its neural network generalizations are that they do not scale well to high dimensions.

In \Ref{frequentstway}, we propose an alternative way to achieve a prior-independent calibration that scales well to high- and variable-dimensional settings.
This approach is based on finding the local maximum likelihood, such that the learned calibration function becomes:
\begin{equation}
\label{eq:ML_calibration}
f_{\rm MLC}(x_D) = \argmax_{z_T}p_{\text{train}}(x_D|z_T)\,,
\end{equation}
where MLC stands for maximum likelihood classifier -- see \Ref{Nachman:2021yvi}.
Again, because the detector response $p(x_D|z_T)$ is universal, maximum likelihood calibrations are universal,%
\footnote{One important caveat is that universality here means prior independence over the space of priors that share the same support as the training set.  One cannot get away with training a model on a single $z_T$ instance and expecting it to work everywhere!}
and in certain configurations, are provably unbiased. In particular, if the detector response $p(x_D|z_T)$ is a Gaussian noise model centered on $z_T$, then one can show that the bias is zero using \Eq{bias}:
\begin{align}
    b(z_T)+z_T&=\int dx_D\,\argmax_{z_T}\left[p(x_D|z_T)\right]\,p(x_D|z_T) \\
    &= \int dx_D\,x_D\,\frac{1}{\sqrt{2\pi \epsilon^2}}e^{-\frac{(x_D-z_T)^2}{2\epsilon^2}} \nonumber \\
    &= z_T. \nonumber
\end{align}
Here, we have made use of the fact that for a Gaussian, $p(x_D|z_T)$ is maximized at $x_D = z_T$, and that the average of this Gaussian is simply $z_T$. 
This conclusion holds even if the detector response includes offsets, or if the noise $\epsilon$ depends on $z_T$.%
\footnote{It is not always true that a maximum likelihood calibration is unbiased. For instance, if $X_D$ is drawn from a uniform distribution $U(0, z_T)$, then the maximum likelihood estimate from a single $x_D$ sample is $\hat{z}_T = x_D$, whereas an unbiased estimate would be $\hat{z}_T = 2x_D$.}

The strategy in \Ref{frequentstway} is to estimate the (local) likelihood density by extremizing the Donsker-Varadhan representation (DVR)~\cite{Donsker1975AsymptoticEO, belghazi2018mine} of the Kullback-Leibler divergence~\cite{kullback1951information}:
\begin{align}\nonumber
L[f] &= \mathbb{E}_{p(x_D,z_T)}\big[f(x_D,z_T)\big] \\
&\quad -\log \mathbb{E}_{p(x_D)p(z_T)}\big[e^{f(x_D,z_T)}\big]\,. \label{eq:DVR_loss}
\end{align}
By parametrizing $f(x_D,z_T)$ via a specially chosen Gaussian Ansatz (see~\Ref{frequentstway} for details), one can extract the local maximum likelihood estimate and resolution with a single neural network training.

We focused on regression in the above discussion, but prior dependence also appears in classification calibration.
A classifier trained with the MSE loss function or the binary cross entropy (BCE) will learn the probability of the signal given an observed $x_D$.
If the fraction of signal is different in the training set and the test set, that is, $p_\text{test}(z_T) \not= p_\text{train}(z_T)$, then the output can no longer be interpreted as the probability of the signal.
Luckily, classifiers are almost never used this way in HEP, since the classification score is not interpreted directly as a probability.%
\footnote{See \Ref{pmlr-v70-guo17a} for a review in the machine learning literature and \Ref{Cranmer:2015bka} for related studies in the context of HEP likelihood ratios.}
In this case, simulation-based calibrations may not be required,%
\footnote{There may be practical issues associated with prior dependence, e.g., if there is an extreme class imbalance, the classifier may not learn well.  In the extreme limit of only one class present in the training, then there is a prior dependence also on the result.}
though data-based calibrations are still essential, as described next.

\subsection{Data-based Calibration}
\label{sec:databased}

In data-based calibration, the goal is to account for possible differences between a true detector response, $p_{\text{data}}(x_D)$ and a simulated detector model $p_{\text{sim}}(x_D)$.
That is, the goal is to match detector level features $x_D$ between data and a simulation at the distribution level, in contrast to simulation-based distribution, where the goal is to match $x_D$ and a target feature $z_T$ at the object level.
Usually, $p_{\text{data}}(x_D)$ is a control dataset, and $p_{\text{sim}}(x_D) = \int dz_T\, p_{\text{sim}}(x_D|z_T)\,p_{\text{train}}(z_T)$ is a simulated detector output generated from truth-level features $z_T$.

In the machine learning literature, data-based calibration is called \textit{domain adaptation}.
Machine learning domain adaptation has been widely studied in the context of HEP~\cite{Rogozhnikov:2016bdp,Andreassen:2019nnm,Cranmer:2015bka,2009.03796,Nachman:2021opi} (see also \textit{decorrelation}~\cite{Louppe:2016ylz,Dolen:2016kst,Moult:2017okx,Stevens:2013dya,Shimmin:2017mfk,Bradshaw:2019ipy,ATL-PHYS-PUB-2018-014,DiscoFever,Xia:2018kgd,Englert:2018cfo,Wunsch:2019qbo,Rogozhnikov:2014zea,10.1088/2632-2153/ab9023,clavijo2020adversarial,Kasieczka:2020pil,Kitouni:2020xgb,Ghosh:2021hrh}), but these tools have not yet been applied to per-object calibrations.
Traditional methods typically use binned or simple parametric approaches to calibrate differences between data and simulation.

The authors of \Ref{Pollard:2021fqv} propose to use tools from the field of \textit{optimal transport} (OT) to perform the data-based calibration using machine learning.
The central idea is to learn a map $h:\mathbb{R}^N\rightarrow\mathbb{R}^N$ that ``moves'' $x_D$ as little as possible, but still achieves $p_\text{sim}(x_D)\mapsto p_\text{data}(x_D)$.
In this case, the OT-based calibration is:
\begin{align}
    \label{eq:OT_detector_calibration}
    \hat{p}(x_D) &= p_{\rm sim}(h(x_D))\,|h'(x_D)|\,,
\end{align}
where $|h'(x_D)|$ is the Jacobian factor.
The precise transportation map depends on the choice of OT metric. 
\Eq{OT_detector_calibration} can be interpreted as shifting simulated samples $x_D$ to $h(x_D)$, and additionally reweighting each sample by $|h'(x_D)|$. 
One can also write a corresponding expression for the OT-calibrated detector model, conditioned on $z_T$:
\begin{align}
    \label{eq:OT_detector_model}
    \hat{p}(x_D|z_T) &= p_{\rm sim}(h(x_D)|z_T) \, |h'(x_D)|\,.
\end{align}

\Eq{OT_detector_model} can be thought of as a ``corrected simulated response'' function that accounts for mismodeling in the original simulation, $p_{\rm sim}(x_D|z_T)$.
At first glance, \Eq{OT_detector_model} might seem prior independent, since it is conditioned on the truth-level $z_T$.
As we will see, though, there is implicit prior dependence in $h$.
For simplicity, consider the special case of one dimension.
Here, for any OT metric, the OT map $h: \mathbb{R} \to \mathbb{R}$ is simply given by:
\begin{equation}
    \label{eq:OT_map_1D}
    h(x_D)=P_\text{data}^{-1}(P_\text{sim}(x_D)),
\end{equation}
where $P_\lambda$ is the cumulative distribution function of $\lambda$, i.e.\ $P_\lambda(x_D)=\int_{-\infty}^{x_D} dx_D'\, p_\lambda(x_D')$.
This function maps quantiles of the simulated distribution to quantiles of the data distribution.
The Jacobian of this transformation is: 
\begin{align}
    |h'(x_D)| &= \frac{p_{\rm sim}(x_D)}{p_{\rm data}(h(x_D))} \\
              &=\frac{\int d z_T \, p_{\rm sim}(x_D|z_T)\,  p_{\rm train}(z_T)}{p_{\rm data}(h(x_D))} \nonumber.
\end{align}
Thus, since the prior $p_{\rm train}(z_T)$ explicitly appears, the derived OT-based detector model in \Eq{OT_detector_model} is prior dependent.

In line with simulation-based calibration, the bias of a data-based calibration is the average difference between the estimator $\hat{p}(x_D)$ and the desired value $p_{\text{data}}(x_D)$, conditioned on $x_T$.%
\footnote{This differs from the simulation-based calibration definition, which was conditioned on $z_T$. In data, there is no truth level $z_T$. However, sometimes, a proxy can be used as a $z_T$ in data, allowing for a direct comparison of true versus reconstructed $z_T$ values in data-based calibration. For example, when performing data-based calibration on a $Z$+jets sample, the $p_T$ of the $Z$ can be used as a proxy for the true jet $p_T$.}
For OT-based calibration, the bias for a given value of $x_D$ is:
\begin{align}
    b(x_D) &= p_{\rm sim}(h(x_D))\,|h'(x_D)| - p_{\rm data}(x_D) \\
           &= \int dz_T\, p_{\rm sim}(h(x_D)|z_T)\,p_{\rm test}(z_T)\,|h'(x_D)| \nonumber \\ &\quad - p_{\rm data}(x_D) \nonumber.
\end{align}
If $p_{\rm test}(z_T) = p_{\rm train}(z_T)$, then the bias is zero. 
Otherwise, the calibration is biased, a consequence of prior dependence. 
Note that this is in contrast to simulation-based calibration, where non-universality can imply a bias even if $p_{\rm test}(z_T) = p_{\rm train}(z_T)$. 

\subsection{Unbiased Data-based Approaches?}
\label{sec:data_based_prior}

As defined above, the goal of a data-based calibration is to match $p_{\rm sim}(x_D)$ to $p_{\rm data}(x_D)$.
This is an inherently prior dependent task, however, since $\hat{p}(x_D) = \int dz_T\, \hat{p}(x_D|z_T)\,p_{\text{train}}(z_T)$ -- that is to say, the simulated detector output depends on the simulation input.
Instead, one can ask if the corrected response function, $\hat{p}(x_D|z_T)$, is universal. 
If it is, then one can use the same corrected response function to generate $\hat{p}(x_D)$ for a variety of priors $p_{\rm test}(z_T)$.  
At least in the special case of one-dimensional OT-based calibration, however, we have shown above that the corrected response function is \emph{not} universal.

To our knowledge, no one has proposed a data-based calibration method that is prior independent, whether using machine learning or not.
This implies that all data-based calibration methods in use are biased, though the degree of bias may be small if the testing and training truth-level densities are similar enough.
We encourage the community to develop a prior-independent data-based calibration strategy, or prove that it is impossible.

\section{Resolution and Uncertainty in Calibrations}\label{sec:resolution_uncertainty}

The discussion thus far has focused on mitigating bias in calibration. Two related concepts are the resolution and uncertainty of a calibration. In this section, we review calibration resolution and uncertainty, and we clarify important nomenclature in HEP settings.

\subsection{Resolution}
\label{sec:resolution}

 As already mentioned, the bias of a calibration refers to the difference in central tendency (such as the mean, median, or mode) between a reconstructed quantity and a reference quantity.
 By contrast, the \emph{resolution} of a calibration refers to the \emph{spread} in the difference between the reconstructed and reference quantities. Using variance as our measure of spread, the resolution $\Sigma^2(z_T)$ can be written as the variance of differences between the reconstructed and truth values, conditioned on the truth values, evaluated over the test sample:
 \begin{align}
     \Sigma^2(z_T) = {\rm Var}_{\rm test}[f(X_D)-z_T|Z_T=z_T].\label{eq:resolution}
 \end{align}

 Resolutions, like biases, can be prior dependent. When using the MSE-based calibration (\Eq{MSE_sol}), this becomes:
 \begin{align}
     &\Sigma^2(z_T) + b^z(z_T) \\ \label{eq:mse_resolution}
     &= \int dx_D\,\left(\int dz_T\,z_T'p_{\rm train}(z_T'|x_D) - z_T\right)^2 p_{\rm test}(x_D|z_T). \nonumber
 \end{align}
The prior dependence is seen by applying Bayes' Theorem to $p_{\rm train}(z_T'|x_D)$.

As before, this prior dependence can be reduced if the prior is wide compared to the detector response. 
If the prior $p(z_T)$ is Gaussian with standard deviation $\sigma$, and the detector response $p(x_D|z_T)$ is a Gaussian noise model with standard deviation $\epsilon$, then by applying \Eq{mse_resolution}, one can show that the resolution scales as:
\begin{align}
    \Sigma^2(z_T)\sim \epsilon^2  + \mathcal{O}\left(\left(\frac{\epsilon}{\sigma}\right)^4\right)\epsilon^2.
\end{align}
 On the other hand, for the prior-independent MLC calibration (\Eq{ML_calibration}), the resolution can be shown to be:
 \begin{align}
     \Sigma^2(z_T) = \epsilon^2 \label{eq:gaussian_resolution}.
 \end{align}

 In HEP (and many other) applications, however, it is common to instead refer to the resolution with respect to a measurement $x_D$ rather than the true value $z_T$. 
 That is, for an inference $\hat{z}_T = f(x_D)$, we would like a measure of the spread of $z_T$ values consistent with this measurement, which we will denote $\Sigma(x_D)$ (distinguished by the $x_D$ argument rather than $z_T$).
 Depending on the context and type of calibration, there are a variety of ways to define $\Sigma(x_D)$ -- for instance, as the standard deviation from a Gaussian fit to the distribution of reconstructed over true energies (see e.g.\ \Ref{1609.05195}).
 For our purposes, we can define the point resolution $\Sigma^2(x_D)$ as the variance of $z_T$'s conditioned on $x_D$:
 \begin{align}
     \Sigma^2(x_D) &=  {\rm Var}_{\rm test}[Z_T|X_D=x_D]\label{eq:point_resolution} \\
     &=  \mathbb{E}_{\rm test}[\left(f_{\rm MSE}(x_D) - Z_T\right)^2|X_D=x_D].\nonumber
 \end{align}
For the MSE-based calibration, this is simply the variance of the posterior, $p(z_T|x_D)$.
However, for frequentist approaches where the posterior is not well defined, such as the maximum likelihood calibration, the resolution cannot be defined this way and care must be taken.
For Gaussian noise models $p(x_D|z_T)$, the likelihood is symmetric under interchanging the arguments $x_D$ and $z_T$, so one can take the resolution to be (applying \Eq{resolution}):
\begin{align}
    \Sigma^2(x_D) &= \Sigma^2(z_T) = \epsilon^2. \label{eq:MLC_point_resolution}
\end{align}

 Calibrations do not necessarily improve the resolution and can sometimes make the resolution seem worse.
 For example, if a calibration requires multiplying the reconstructed quantity by a fixed number greater than one, then the resolution will grow by the same amount.%
 \footnote{This is also true if we had used the relative resolution, $\mathbb{E}\left[\frac{f(x_D)}{z_T} | Z_T = z_T\right]$, which is also commonly used in HEP, rather than the absolute resolution.}
 It is therefore important to compare resolutions only after calibration.

 If a calibration incorporates many features that determine the resolution of a given quantity, then the resolution can improve from calibration.
 For example, suppose the reconstructed value $x_D$ is some function of observable quantities $\Vec{y}_D = (y_{D1}, y_{D2}, ..., y_{Dn})$, i.e.\ $x_D = g(\Vec{y}_D)$.
 For instance, in the context of jet energy calibrations, $x_D = \alpha \, \eta$ for some constant $\alpha$ and an observable quantity $\eta$ (e.g.\ energy dependence on the pseudorapidity).
 If any of the $\Vec{y}_D$ have a non-trivial probability density, this will be inherited by the reconstructed value $x_D$ and thus $x_D$ will have a non-zero resolution.
 This resolution is completely reducible, however, through a calibration that is $\Vec{y}_D$ dependent -- that is, a calibration function $\hat{z}_T = f'(\Vec{y}_D)$ rather than $\hat{z}_T = f(x_D)$.
 The ability to incorporate many auxiliary features is why machine-learning-based approaches, such as the Gaussian Ansatz~\cite{frequentstway}, have the potential to improve analyses at HEP experiments.

\subsection{Uncertainty}
\label{sec:uncert}

In the machine learning literature, ``resolution'' would be referred to as a type of ``uncertainty''.
Uncertainty in the statistical context refers to the limited information about $z_T$ contained in $x_D$.
In the HEP literature, though, we use uncertainty in a different way, to instead refer to the limited information we have about the bias and resolution of a calibration.

The reason for this difference in nomenclature is that HEP research is based primarily on simulation-based inference, where data are analyzed by comparison to model predictions.
(This is the case for the vast majority of analyses at the LHC.)
In this context, the word ``uncertainty'' is reserved to refer to uncertainties on model parameters. 
A worse resolution can degrade the statistical precision of a measurement, but if it is well modeled by the simulation, then there is no associated systematic uncertainty (though there will still be statistical uncertainties).

Both simulation-based and data-based calibrations can have associated uncertainties.
For simulation-based calibrations, even if they are prior independent, there can be uncertainties in the detector models themselves.
For data-based calibrations, there are additional uncertainties associated with the truth-level prior; see \Sec{data_based_prior}.

One of the goals of data-based calibration is to improve the modeling of the calibration in simulation to match the data.
Typically, data-based calibrations are performed in dedicated event samples with well-understood physics processes.
The residual uncertainty following the data-based calibration is dominated by the modeling of the underlying process.
For example, data-based jet calibrations (called ``in situ'' calibrations) compare the jet to a well-measured reference object such as a $Z$ boson.
The momentum imbalance between the jet and the $Z$ boson will be due in part to differences in the calibration between data and simulation and in part due to the mismodeling of initial and final state radiation.
Uncertainties on the latter are then incorporated into the data-based calibration uncertainty.
In nearly all cases, data-based calibrations are performed independent of the uncertainties, which are computed post-hoc.
In the future, these uncertainties may be improved with uncertainty/inference-aware machine learning methods~\cite{Blance:2019ibf,Englert:2018cfo,Louppe:2016ylz,Dolen:2016kst,Moult:2017okx,Stevens:2013dya,Shimmin:2017mfk,Bradshaw:2019ipy,ATL-PHYS-PUB-2018-014,DiscoFever,Wunsch:2019qbo,Rogozhnikov:2014zea,10.1088/2632-2153/ab9023,clavijo2020adversarial,Kasieczka:2020pil,Kitouni:2020xgb,Estrade:2019gzk,Wunsch:2020iuh,Elwood:2020pik,Xia:2018kgd,deCastro:2018mgh,Charnock_2018,Alsing:2019dvb,lukas_heinrich_2020_3697981,Kasieczka:2020vlh,Bollweg:2019skg,Araz:2021wqm,Bellagente:2021yyh,Nachman:2019dol,1807719,Ghosh:2021hrh,Ghosh:2021roe}.

\section{Gaussian Examples}
\label{sec:gaussian}

In this section, we demonstrate some of the calibration issues related to bias and prior dependence in a simple Gaussian example.
We assume that the truth information (the ``prior'') is distributed according to a Gaussian distribution with mean $\mu$ and variance $\sigma^2$:
\begin{equation}
Z_T\sim\mathcal{N}(\mu,\sigma^2).
\end{equation}
The detector response is assumed to induce Gaussian smearing centered on the truth input with variance $\epsilon^2$:
\begin{equation}
X_D|Z_T=z_T\sim\mathcal{N}(z_T,\epsilon^2).
\end{equation}

For the simulation-based calibration in \Sec{simGauss}, the goal is to learn $Z_T$ given $X_D$, assuming perfect knowledge of the detector response.
For the data-based calibration in \Sec{dataGauss}, the goal is to map $X_D$ in ``simulation'' to $X_D$ in ``data''.
In this latter study, we assume that data and simulation have the same true probability density and differ only in their detector response, $\epsilon_\text{sim}\neq \epsilon_\text{data}$ -- that is, $p_{\rm sim}$  ``mismodels'' $p_{\rm data}$. 

\subsection{Simulation-based Calibration}
\label{sec:simGauss}

If we use the MSE approach in \Eq{MSE_sol}, there is a prior dependence in the calibration, which induces bias.
Perhaps counter-intuitively, this bias persists even if the prior is the same as the data density:
\begin{equation}
\label{eq:equal_prior}
p_\text{train}=p_\text{test}\equiv p,
\end{equation}
as we now show.

\begin{figure*}[t]
    \centering
    \subfloat[]{
         \includegraphics[width=0.415\textwidth]{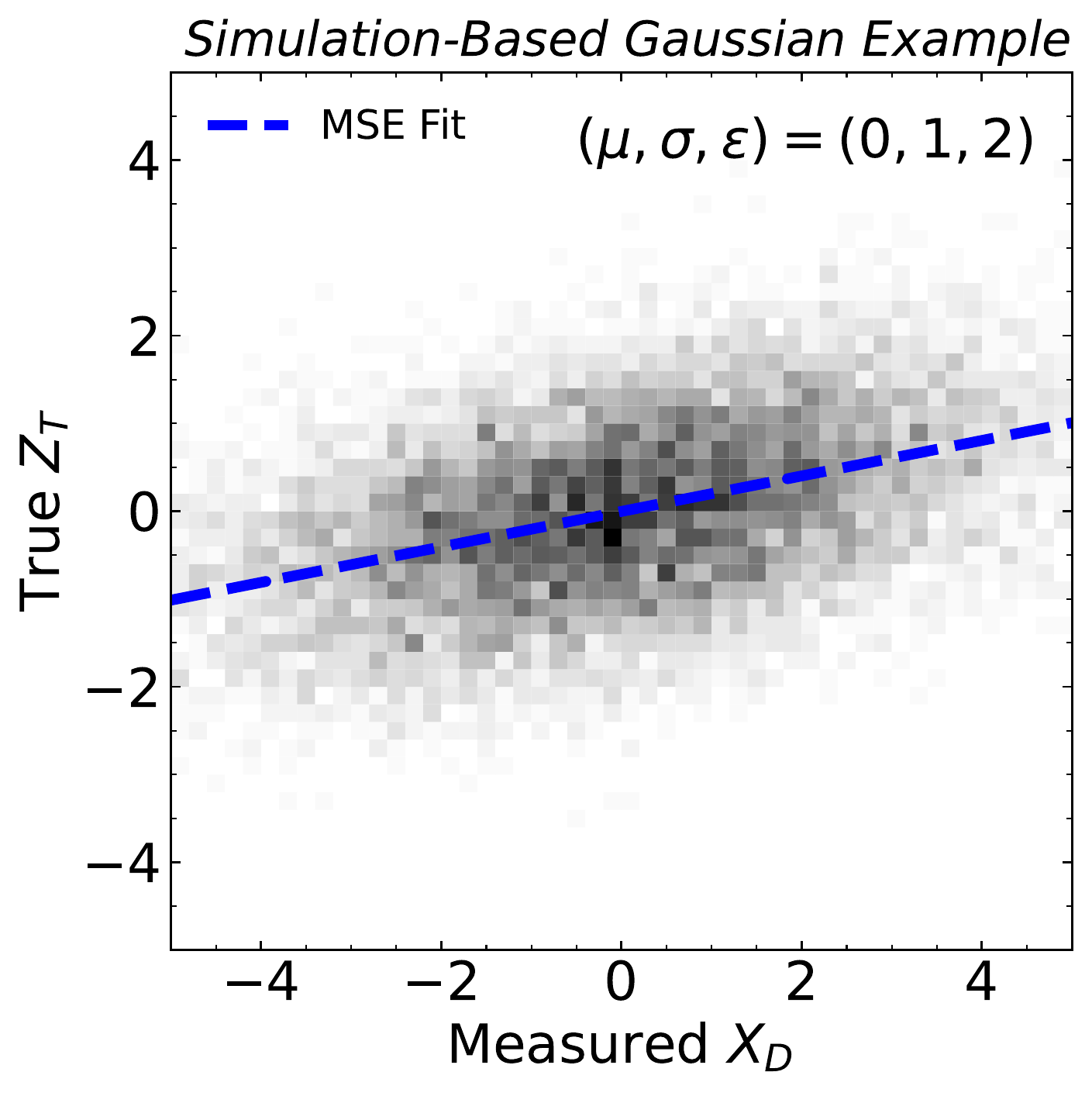}
        \label{fig:gauss1a}
    }
    $\qquad$
    \subfloat[]{
        \includegraphics[width=0.415\textwidth]{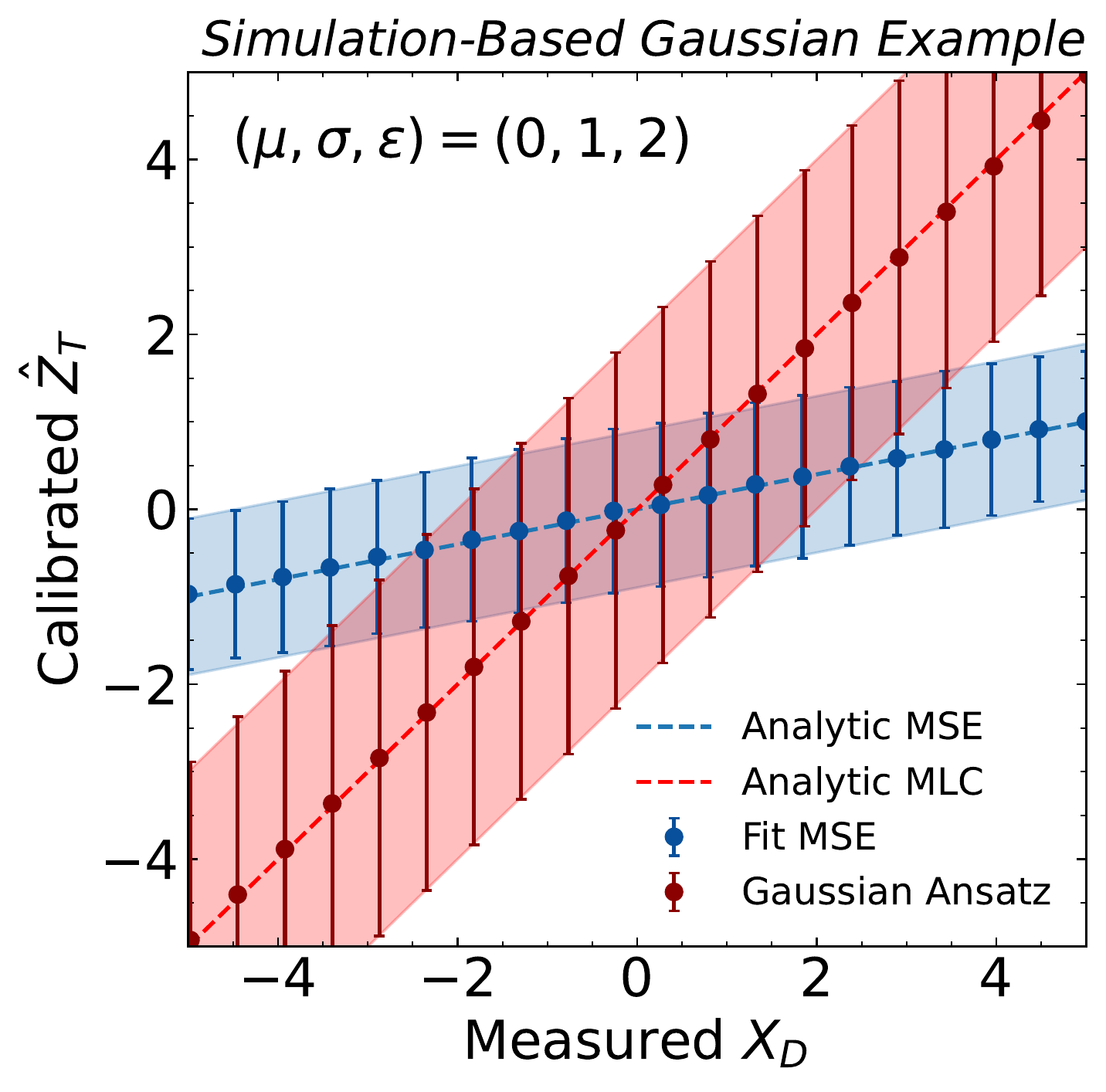}
        \label{fig:gauss1b}
    }
    \caption{
        (a) 
        2D Histogram of the reconstructed value $x_D$ distribution versus the true value $z_T$ distribution, in the Gaussian example with $\mu=0$, $\sigma=1$, and $\epsilon=2$.
        The dashed line represents a linear fit to the data points.
        (b)
        For test values of $x_D$, the vertical axis is the calibrated target value $ \hat{z}_T(x_D)$.
        The blue dots are the results from a numerical MSE fit $f_{\rm MSE}(x_D)$, and the error bars correspond to the numerical point resolution $\Sigma_{\rm MSE}(x_D)$, with the analytic prediction in the red dotted line.
        For comparison, the Gaussian Ansatz calibration is indicated by the red points $f_{\rm MLC}(x_D)$, with the error bars indicating the point resolution $\Sigma_{\rm MLC}(x_D)$.
        For both fits, the colored lines and bands are the analytically expected results for the fits and resolutions, respectively.
        }
    \label{fig:gauss1}
\end{figure*}

In the Gaussian case, the reconstructed data are distributed according to:
\begin{equation}
    X_D\sim\mathcal{N}(\mu,\sigma^2+\epsilon^2),
\end{equation}
and it is possible to solve \Eq{MSE_fit} analytically, in the asymptotic limit:
\begin{align}
    f_{\rm MSE}(x_D) = \frac{\epsilon^2\mu + \sigma^2 x_D}{\epsilon^2 + \sigma^2}.
\end{align}
For comparison, we can also compute the unbiased maximum likelihood calibration using \Eq{ML_calibration}:
\begin{align}
    f_{\rm MLC}(x_D) = x_D.
\end{align}
It is also possible to analytically compute the point resolutions, $\Sigma_(x_D)$, for both the MSE and MLC fits (\Eqs{point_resolution}{MLC_point_resolution}, respectively):
\begin{align}
    \Sigma_{\rm MSE}(x_D) &= \frac{\epsilon\sigma}{\sqrt{\epsilon^2 + \sigma^2}},  \\
    \Sigma_{\rm MLC}(x_D) &=  \epsilon.
\end{align}

To illustrate this setup, we simulate this scenario numerically for $\mu=0$, $\sigma=1$, and $\epsilon=2$.
In \Fig{gauss1a}, we show the simulated data, for which both the true and reconstructed values follow a Gaussian distribution.
The first step of a typical calibration is to predict the true $z_T$ from the reconstructed $x_D$.
Since we know that the average dependence of the true $z_T$ on the reconstructed $x_D$ is linear, we perform a first-order polynomial fit to the data using \texttt{numpy polyfit}, which is represented by the blue dashed line in \Fig{gauss1a}.
This calibration function is then applied to all reconstructed values:
\begin{equation}
    \hat{z}_T(x_D) = f_{\rm MSE}(x_D)\,.
\end{equation}
The resulting calibration curve is presented in blue in \Fig{gauss1b}, along with the associated resolution $\Sigma_{\rm MSE}(x_D)$.

\begin{figure*}[t]
    \centering
    \subfloat[]{
        \includegraphics[width=0.415\textwidth]{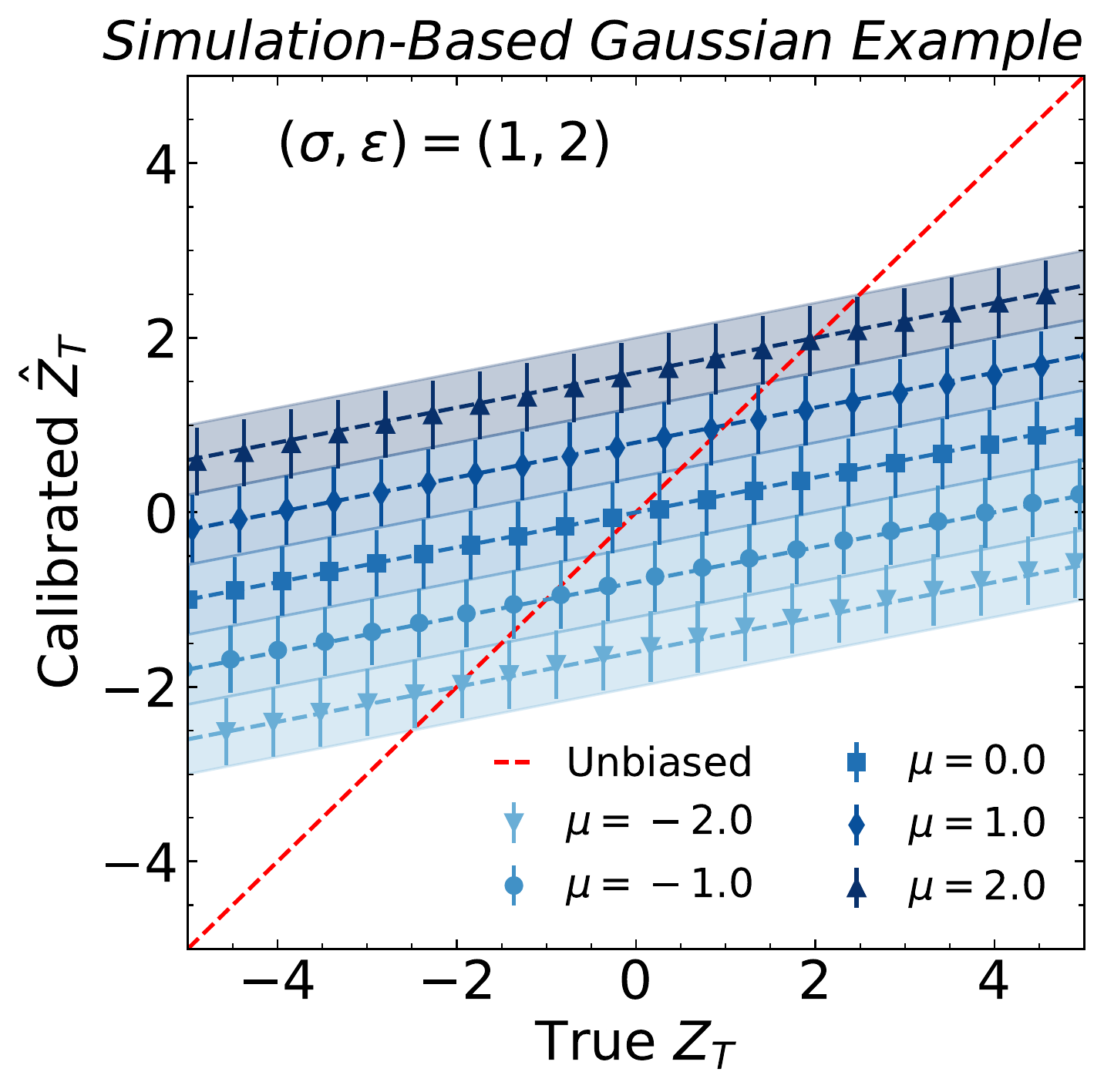}
        \label{fig:gauss2a}
    }
    $\qquad$
    \subfloat[]{
        \includegraphics[width=0.415\textwidth]{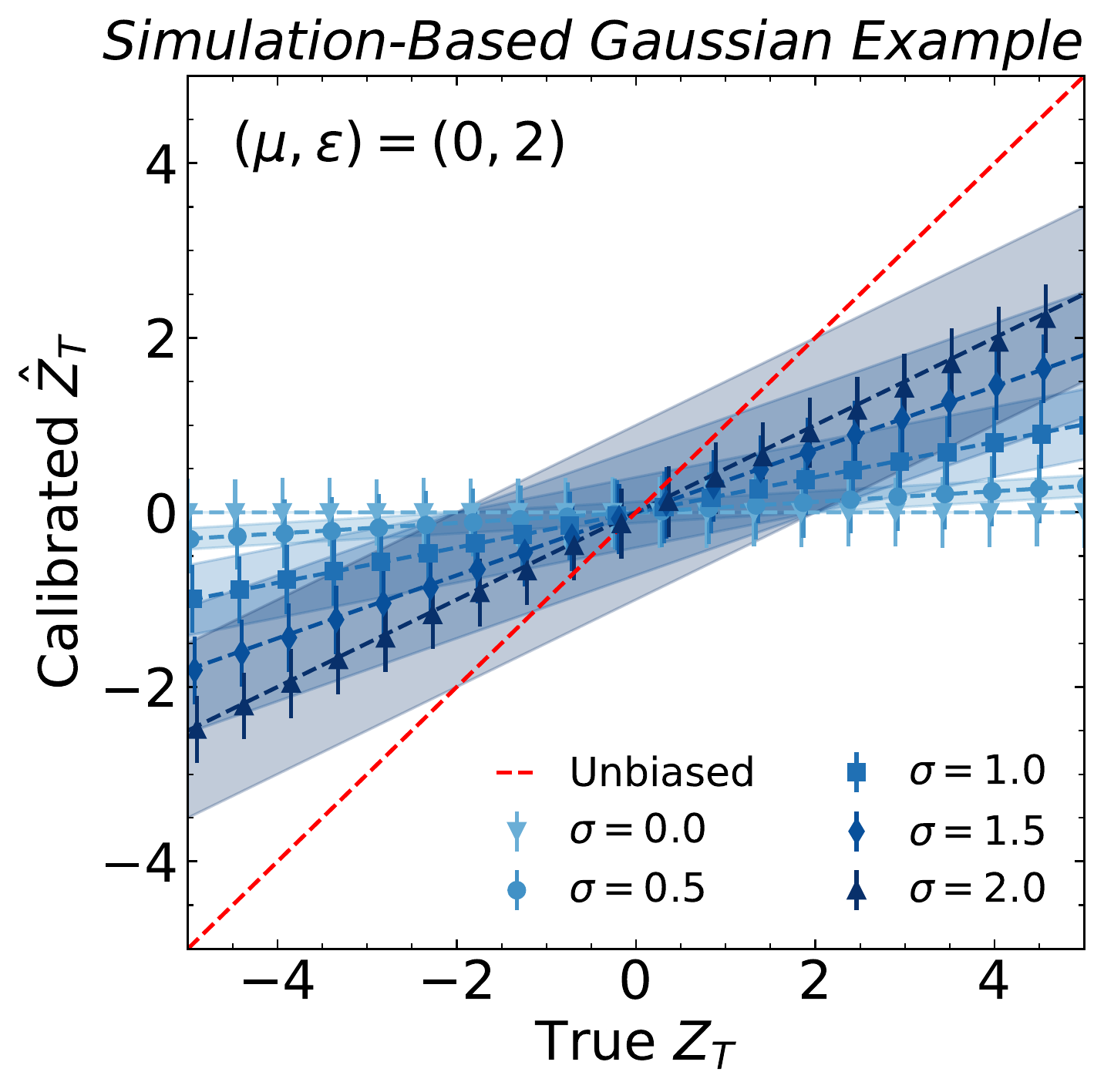}
        \label{fig:gauss2b}
    }
    \caption{The same MSE results as \Fig{gauss1b}, but plotted in bins of true $z_T$ rather than $x_D$. Points correspond to numerical fit results with associated resolution $\Sigma_{\rm MSE}(z_T)$, while the dashed lines and bands correspond to analytic results. Multiple values of the prior parameters (a) $\mu$ and (b) $\sigma$ are shown to illustrate the prior dependence of the bias.  Though not shown, we verified that the Gaussian Ansatz gives results consistent with the unbiased calibration in dashed red.}
    \label{fig:gauss2}
\end{figure*}

For comparison, we perform a maximum likelihood calibration using the Gaussian Ansatz introduced in~\Ref{frequentstway}:
\begin{align}
    f(x,z) &= A(x) + \big(z-B(x)\big)\cdot D(x) \nonumber\\&\quad + \frac{1}{2} \big(z-B(x)\big)^T \cdot C(x,z) \cdot \big(z-B(x)\big)\,,
    \label{eq:gaussian_ansatz}
\end{align}
where we have dropped the subscripts ($x_D \to x$, $z_T \to z$) for compactness of notation.
As described in \Ref{frequentstway}, the calibration function $B(x)$ is obtained by minimizing the DVR loss function from \Eq{DVR_loss}, such that after training:
\begin{align}
    \hat{z}_T(x_D) &= B(x_D),\\
    \Sigma_{\rm MLC}(x_D) &= - \big[C(x_D,B(x_D))\big]^{-1/2}.
\end{align}
For Gaussian noise models, this maximum likelihood estimate is unbiased, as confirmed by the numerical results in \Fig{gauss1b}.
We implement the Gaussian Ansatz in \textsc{Keras}~\cite{keras} with the \textsc{Tensorflow} backend~\cite{tensorflow}.
The $A$ network consists of three hidden layers with 16 nodes per layer, with rectified linear unit activations.
The $B$ and $C$ networks are each a single node with linear activation.
The $D$ network is set to zero by hand.
Optimization is carried out with \textsc{Adam}~\cite{adam} over 100 epochs with a batch size of 128.
As desired, the Gaussian Ansatz yields a calibration that is independent of the prior $p_{\text{train}}(z_T)$.

To demonstrate the bias, we plug in \Eq{equal_prior} into \Eq{closure} to get the bias from the MSE calibration approach:
\begin{equation}
\label{eq:closuregaussian}
b(z_T) + z_T =\int dx_D\,dz_T'\,z_T'\,p(x_D|z_T')\, p(x_D|z_T)\,\frac{p(z_T')}{p(x_D)}.
\end{equation}
It is possible to solve \Eq{closuregaussian} analytically for the Gaussian setup:
\begin{align}
\label{eq:closuregaussian2}
b(z_T) = \left(\frac{\epsilon^2}{\sigma^2+\epsilon^2}\right)(\mu - z_T).
\end{align}
As expected, $b(z_T)\rightarrow 0$ as $\epsilon\rightarrow 0$.
For $\epsilon>0$, though, there is a non-zero bias with the MSE approach. 
The $z_T$-binned resolutions can also be computed using \Eqs{mse_resolution}{gaussian_resolution}:
\begin{align}
    \Sigma_{\rm MSE}(z_D) &= \frac{\sigma^2}{\epsilon^2 + \sigma^2} \epsilon, \\
    \Sigma_{\rm MLC}(z_D) &=  \epsilon.
\end{align}

The fitted biases and resolutions are presented in \Fig{gauss2}, which exhibits the bias expected from \Eq{closuregaussian2}.
This illustrates the large bias introduced by the MSE regression procedure.

To further highlight the role of prior dependence, we repeat the MSE calibration procedure, where we test multiple values of the prior parameters $\mu$ and $\sigma$ to confirm the predictions in \Eq{closuregaussian2}.
As shown in \Fig{gauss2a}, changes in $\mu$ simply shift the calibration up and down, but do not improve the calibration quality across the true values of $z_T$.
As shown in \Fig{gauss2b}, changes in $\sigma$ change the slope of the calibration.
In the limit $\sigma\rightarrow \infty$, the calibration curve approaches the unbiased curve, as anticipated from \Eq{bias_scaling}.

\subsection{Data-based Calibration}
\label{sec:dataGauss}

As discussed in \Sec{data_based_prior}, we are unaware of any prior-independent data-based calibration.
To highlight this challenge, we study the OT-based technique introduced in \Ref{Pollard:2021fqv} and mentioned in \Sec{databased}.
In our Gaussian example, the goal is to calibrate a ``simulation'' sample with $(\mu_\text{sim.},\sigma_\text{sim.},\epsilon_\text{sim.})$ to match a ``data'' sample with $(\mu_\text{data},\sigma_\text{data},\epsilon_\text{data})$.

\begin{figure*}[t]
    \centering
    \subfloat[]{\includegraphics[width=0.415\textwidth]{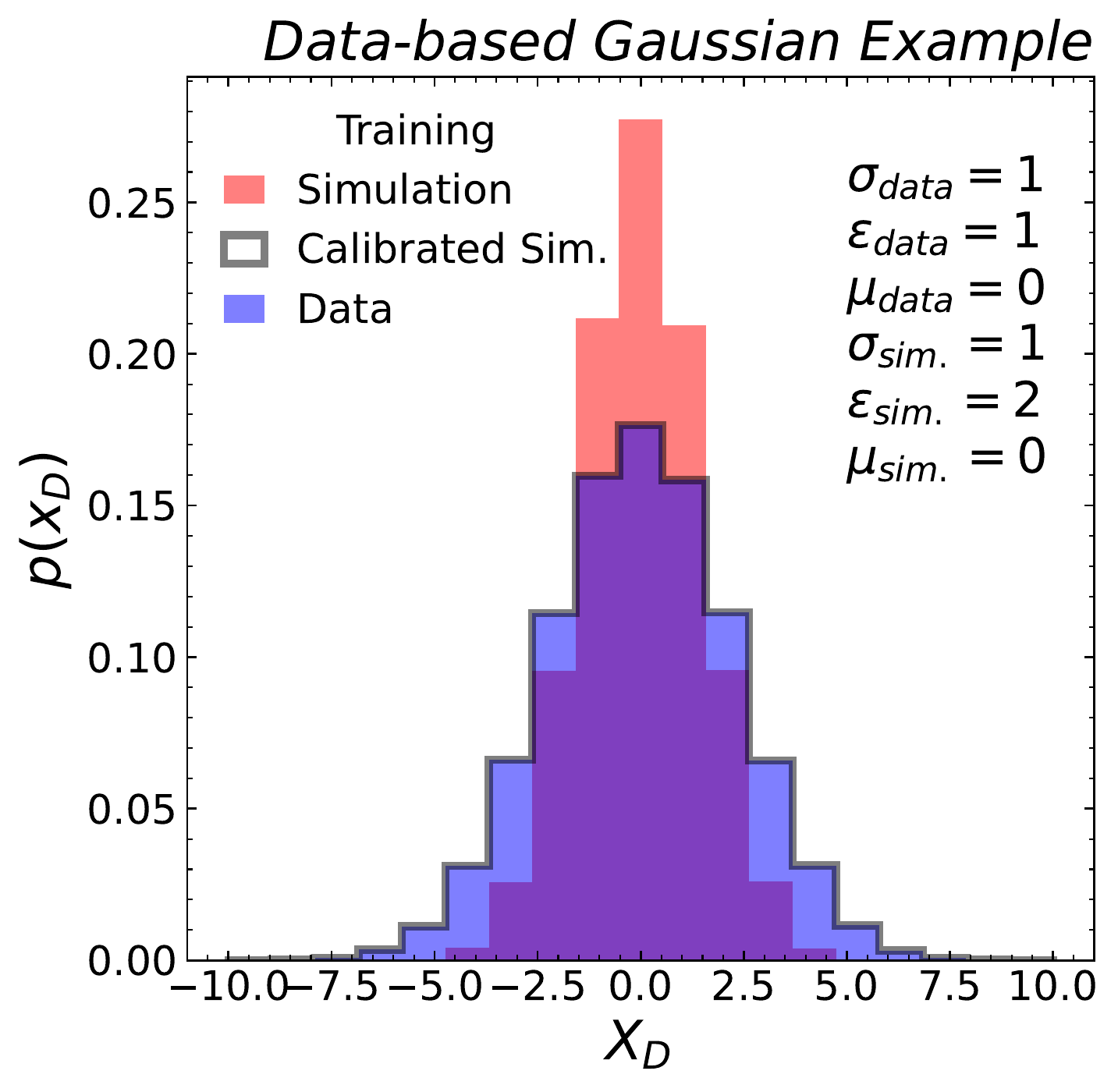}  \label{fig:gauss3a}}
    $\qquad$
    \subfloat[]{\includegraphics[width=0.415\textwidth]{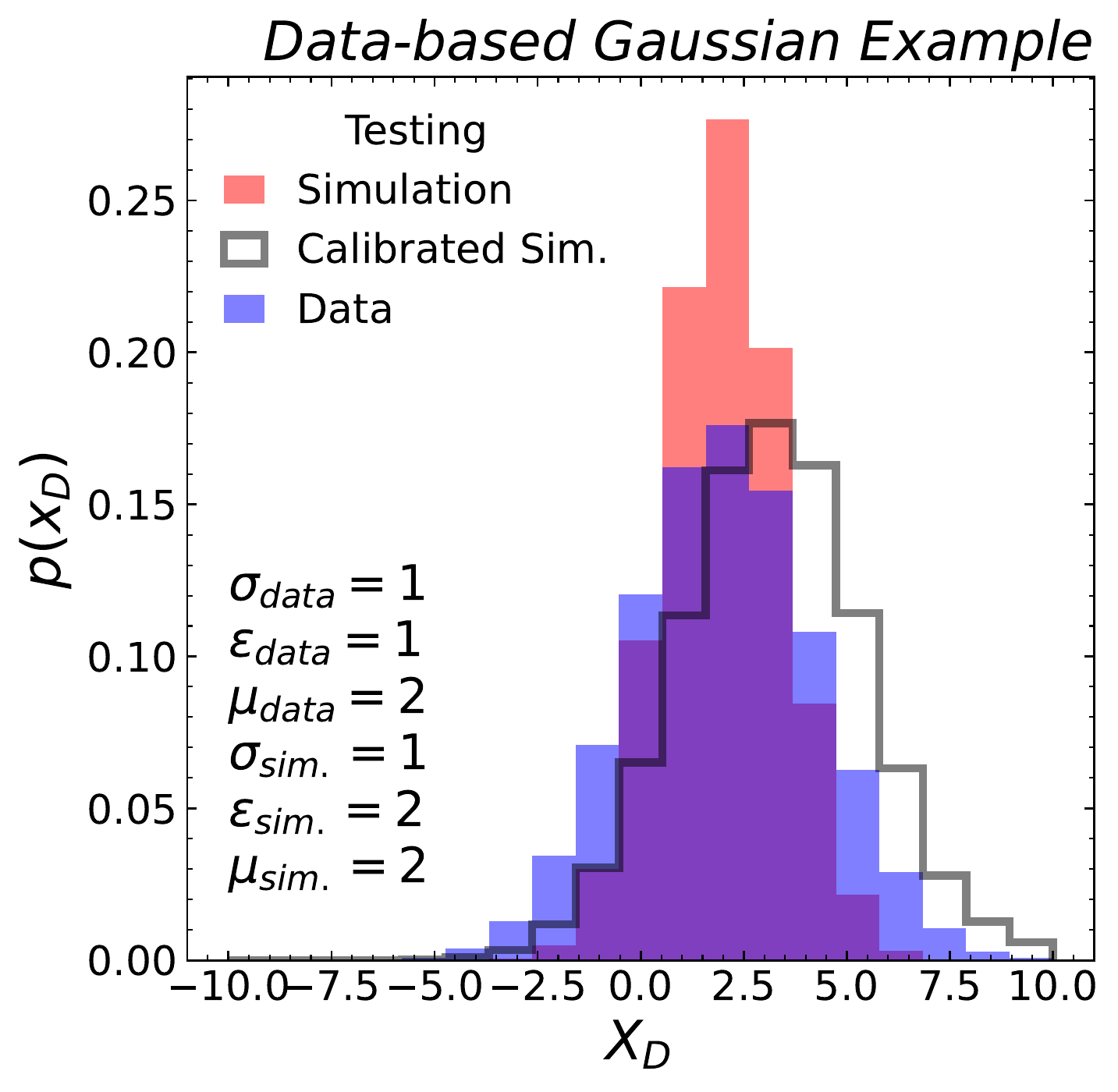}  \label{fig:gauss3b}}
    \caption{
    Histograms of the raw simulation, calibrated simulation, and data for (a) the training set and (b) the test set, the Gaussian example of data-based calibration.
    The calibration function for the test set is used in both figures.}
    \label{fig:gauss3}
\end{figure*}

For simplicity, we assume that the true spectra (determined by $(\mu,\sigma)$) are the same in data and in simulation, such that there is no systematic uncertainty in the calibration (see \Sec{uncert}). 
Only $\epsilon$, the parameter governing the detector response, is different between simulation and data -- the simulation mismodels the real detector.
To highlight the issue of prior dependence, we consider a ``training'' set with one value of $\mu_{\rm train} = 0$ and a ``testing'' set with a different value of $\mu_{\rm test}$, with a shared value of $\sigma$.
The calibration will be derived on the training set and deployed on the testing set.
Again for simplicity, we assume that detector effects (determined by $\epsilon$) are the same in both the train and test sets.

The one-dimensional OT map $h$ from one Gaussian $A$ to another Gaussian $B$ can be computed analytically:
\begin{align}
\label{eq:OTgaussian}
    h_{A\rightarrow B}(x)=\frac{x-\mu_A}{\sigma_A}\cdot\sigma_B+\mu_B,
\end{align}
where the mean and standard deviation of sample $i$ are $\mu_i$ and $\sigma_i$, respectively.
This equation can be derived following \Eq{OT_map_1D}, by computing cumulative distribution function (CDF) of sample $A$ with the inverse CDF of sample $B$.

For the training set with $\mu_{\rm train} = 0$, we have
\begin{align}
    \label{eq:testset}
    h_\text{train}(x) &= \frac{\sqrt{\sigma^2+\epsilon_\text{data}^2}}{\sqrt{\sigma^2+\epsilon_\text{sim}^2}} \, x \\ \nonumber
    &\equiv \alpha \, x.
\end{align}
The test set only differs in the value of $\mu_{\rm test}$, so the correct calibration function should be:
\begin{align}
\label{eq:trainset}
    h_\text{test}(x)&=\alpha(x-\mu_{\rm test})+\mu_{\rm test}\\
    \nonumber
    &=\alpha x + \mu_{\rm test}(1-\alpha).
\end{align}
As long as $\alpha\neq 1$, then $h_\text{train}\neq h_\text{test}$ and so the calibration is not universal.

\begin{figure}[t]
    \centering
    \includegraphics[width=0.415\textwidth]{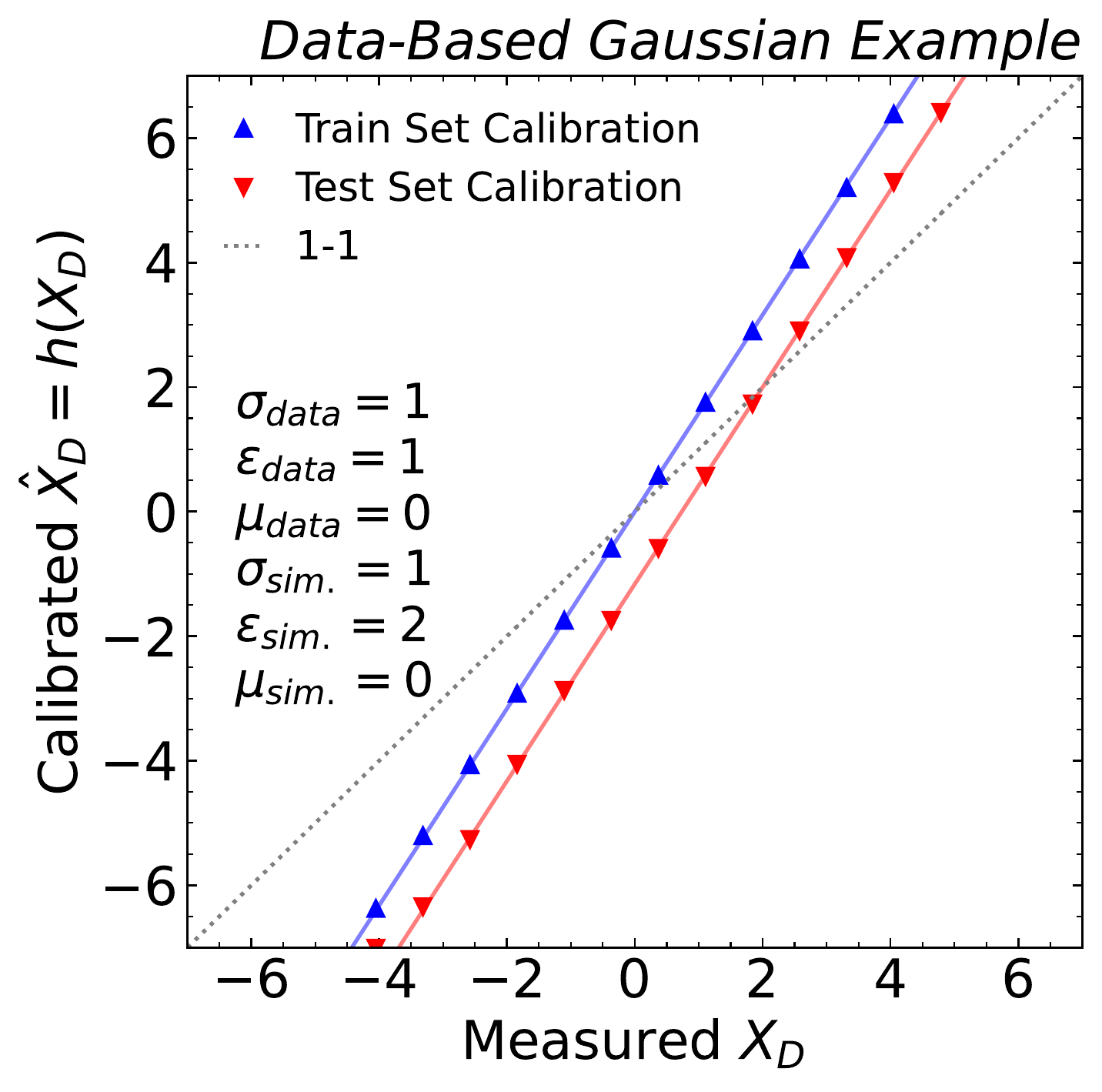}
    \caption{
    The data-driven calibration functions corresponding to \Fig{gauss3}.
    The blue points correspond to the calibration function $h_{\rm train}$ derived from the training set and the red points correspond to the ideal calibration $h_{\rm test}$ one would derive from the test set.}
    \label{fig:gauss4}
\end{figure}

A numerical demonstration of this bias is presented in \Fig{gauss3}, where histograms of the data and simulation are presented along with the calibrated result.
In \Fig{gauss3a}, we see the calibration derived in the training sample, where by construction, the calibrated simulation matches the data.
Since the truth distribution is different in the test set, however,
the training calibration applied in the test set is biased, as shown in \Fig{gauss3b}.
The actual calibration function is plotted in \Fig{gauss4} and compared to the analytic expectation from \Eqs{trainset}{testset}.
The fact that the calibration derived on the train set is not the same as the calibration derived on the test set shows that the calibration derived in one and applied to the other will lead to a residual bias.

\section{Calibrating Jet Energy Response}
\label{sec:hep}

Jets are ubiquitous at the LHC, and their calibration is an essential input to a majority of physics analyses performed by ATLAS and CMS.
In this section, we consider a simplified version of simulation-based and data-based jet energy calibrations.
To illustrate the impact of the prior dependence, we use a realistic and also extreme example where calibrations are derived in a sample of generic quark and gluon jets and then applied to a test sample of jets from the decay of a heavy new resonance.
To further simplify the problem, we consider a calibration of the invariant mass $m_{jj}$ of the leading two jets.
In practice, jet energy calibrations are derived for individual jets, but this requires at least including calibrating the jet rapidity in addition to the jet energy.
We keep the problem one-dimensional in order to ensure the problem is easy to visualize and to mitigate the dependence on features that are not explicitly modeled.
For a high-dimensional study of jet energy calibrations in a prior-independent way, see~\Ref{frequentstway}.

\subsection{Datasets}

Our study is based on generic dijet production in quantum chromodynamics (QCD).
For these studies, we will consider two different datasets to demonstrate simulation-based and data-based jet energy calibrations.
The first dataset is made with a full detector simulation. 
The full simulation sample uses \textsc{Pythia}~6.426~\cite{Sjostrand:2006za} with the Z2 tune~\cite{Chatrchyan:2011id} and interfaced with a \textsc{Geant4}-based~\cite{Agostinelli:2002hh,1610988,Allison:2016lfl} full simulation of the CMS experiment~\cite{Chatrchyan:2008aa}.
In simulation-based calibration, our goal will be to reconstruct the truth-level $z_T = m^{\text{true}}_{jj}$ from the detector-level $x_D = m^{\text{reco}}_{jj}$.
The second dataset is constructed with a fast detector simulation.
The fast simulation uses \textsc{Pythia}~8.219~\cite{Sjostrand:2007gs} interfaced with \textsc{Delphes}~3.4.1~\cite{deFavereau:2013fsa,Mertens:2015kba,Selvaggi:2014mya} using the default CMS detector card.
In data-based calibration, our goal will be to match this fast simulation to ``data'', which will be represented by the full simulation.
The full simulation sample comes from the CMS Open Data Portal~\cite{CMS:QCDsim1000-1400,CMS:QCDsim1400-1800,CMS:QCDsim1800} and processed into an MIT Open Data format~\cite{Komiske:2019jim,komiske_patrick_2019_3341502,komiske_patrick_2019_3341770,komiske_patrick_2019_3341772}.
The fast simulation sample is available at \Ref{2107.08979,nachman_benjamin_2021_5108967}.

For each dataset, we have access to the parton-level hard-scattering scale $\hat{p}_T$ from \textsc{Pythia}, which is in general different from the jet-level transverse momentum $p_T$ we are interested in studying.
To avoid any issues related to the trigger, we focus on events where $\hat{p}_T> 1$~TeV.
Particles (at truth level) or particle flow candidates (at reconstructed level) are used as inputs to jet clustering, implemented using \textsc{FastJet}~3.2.1~\cite{Cacciari:2011ma,Cacciari:2005hq} and the anti-$k_t$ algorithm~\cite{Cacciari:2008gp} with radius parameter $R=0.5$.
No calibrations are applied to the reconstructed jets.  

\begin{figure}[t]
    \centering
    \includegraphics[width=0.445\textwidth]{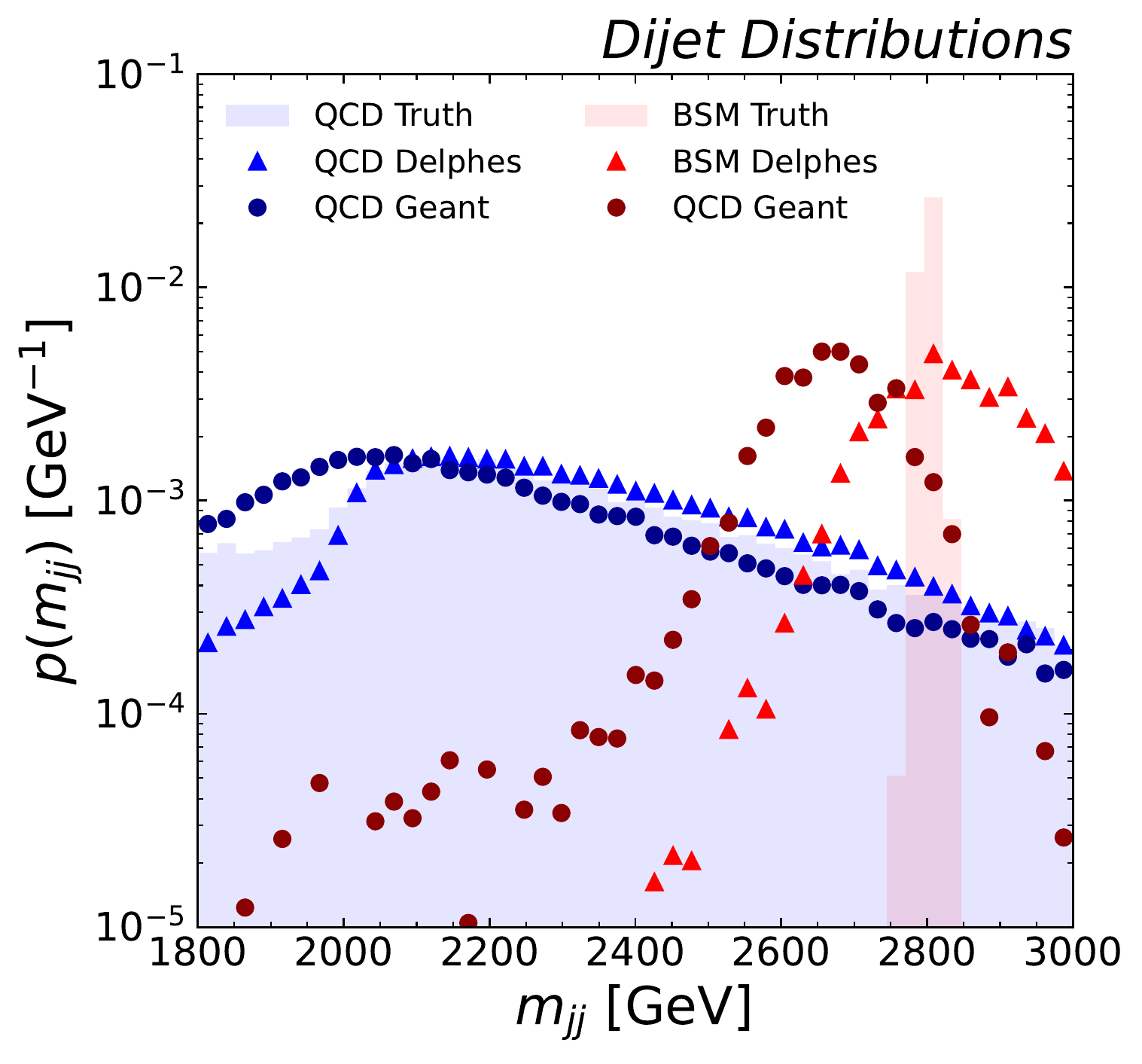}
    \caption{
    The $m_{jj}$ distributions for QCD (blue) and BSM (red) events in the fast and full simulation. The shaded histograms correspond to the $z_T = m^{\text{true}}_{jj}$ truth-level distributions, whereas the light triangles and dark circles correspond to $x_D = m^{\text{reco}}_{jj}$ for the fast (\textsc{Delphes}) and slow (\textsc{Geant4}) distributions respectively.}
    \label{fig:QCD_dijets}
\end{figure}

To emulate two different physics processes while controlling for all hidden variables, we consider dijet events with two different sets of event weights. 
This will allow us to study the prior-dependent effects of each calibration.
\begin{itemize}
    \item \textbf{QCD}.
    This set of weights $\{w_i\}$ comes from the original \textsc{Pythia} event generation.
    The resulting spectra are steeply falling in the invariant mass of the two jets, $m_{jj}$. 
    \item \textbf{BSM}.
    To emulate a narrow dijet resonance, we consider a second set of weights given by
    \begin{equation}
        w(m^\text{true}_{jj,i})\propto \frac{1}{\sigma w_i} \exp\left[-\frac{(m^\text{true}_{jj,i}-\mu)^2}{2\sigma^2} \right],
    \end{equation}
    where $\mu=2.8$ TeV and $\sigma=10$ GeV.
    Note that the weighting is applied using the true $m_{jj}$.
\end{itemize}
The $m_{jj}$ distributions as described above are shown in \Fig{QCD_dijets}.
In the full simulation, one can see a difference between $m^{\text{true}}_{jj}$ and $m^{\text{reco}}_{jj}$ for both QCD and BSM, necessitating a simulation-based calibration.
Additionally, the $m^{\text{reco}}_{jj}$ distribution is significantly different between the full and fast simulations, which to correct requires a data-based calibration.

\begin{figure*}[t]
    \centering
    \subfloat[]{
        \includegraphics[width=0.445\textwidth]{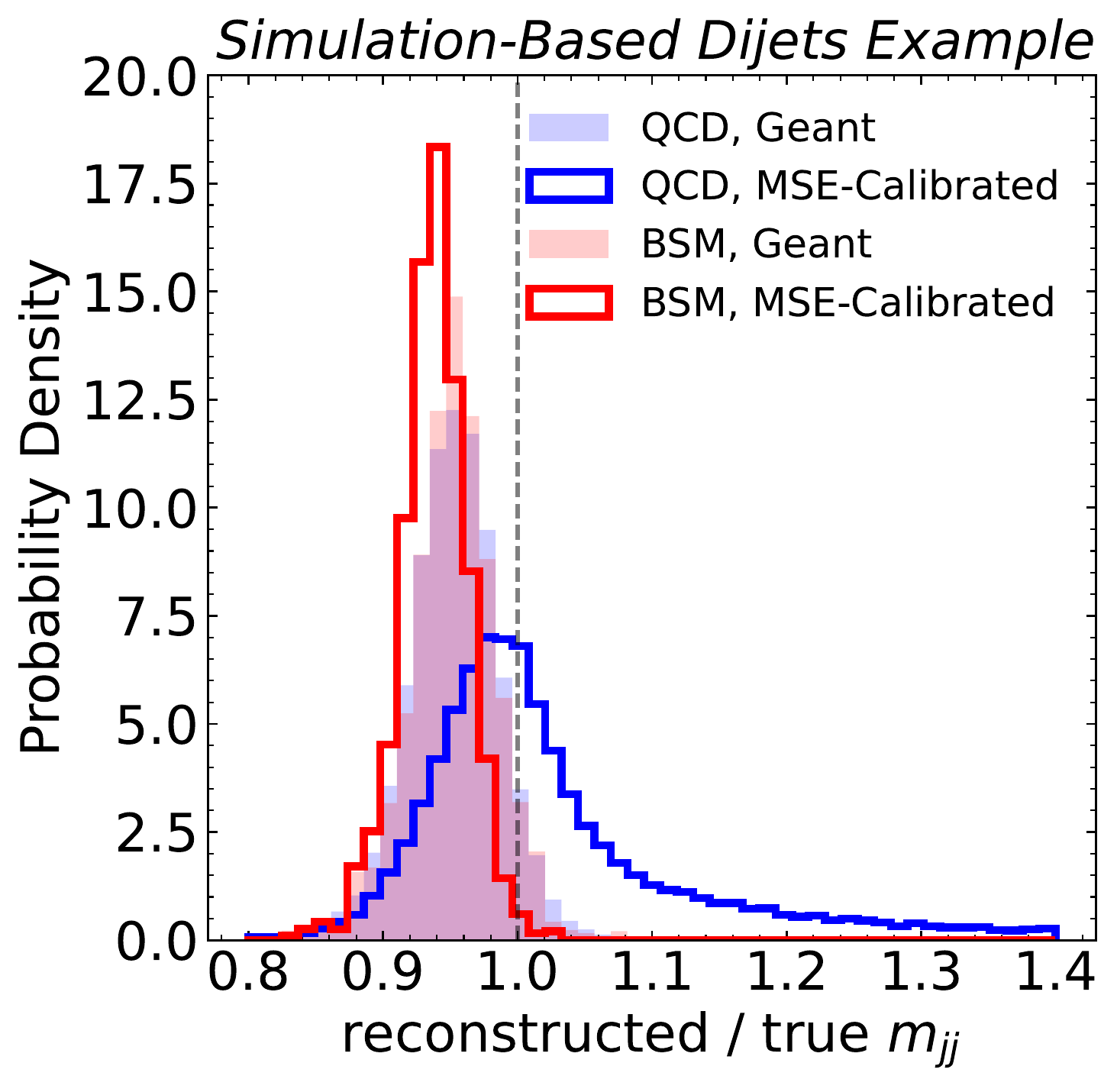}
        \label{fig:MSE_fit}
    }
    $\qquad$
    \subfloat[]{
        \includegraphics[width=0.445\textwidth]{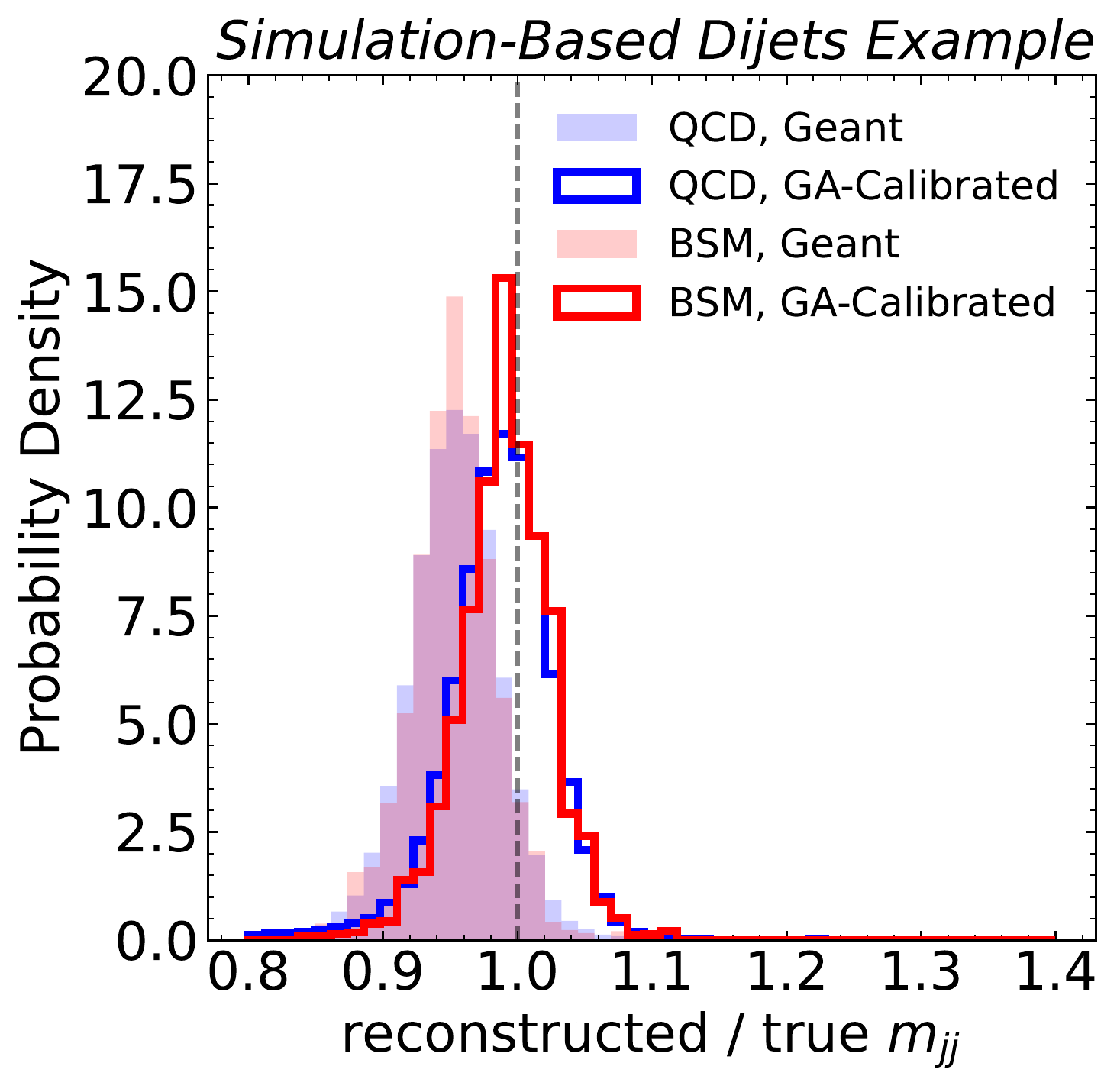}
        \label{fig:GA_fit}
    }
    \caption{
    The reconstructed $m_{jj}$ divided by the true $m_{jj}$ for the QCD and BSM samples, using (a) the MSE-based approach and (b) the maximum likelihood approach with the Gaussian Ansatz.
    Shown are results with and without the simulation-based calibration applied.
    }
\end{figure*}

For all following results, half of the examples are used for training and half are used for testing.

\subsection{Simulation-based Calibration}

The goal for the simulation-based calibration is to learn a function to predict $z_T = m^{\text{true}}_{jj}$ from $x_D = m^{\text{reco}}_{jj}$ in the full simulation.
In contrast to the Gaussian example in \Sec{simGauss}, we do not know the functional form of the calibration.
Therefore, we use a neural network to provide a flexible parametrization of the calibration and numerically minimize the MSE loss.
The neural network has three hidden layers with 50 nodes per layer, with the rectified linear unit activation for intermediate layers and a linear activation for the output.
The network is implemented in \textsc{Keras} with the \textsc{Tensorflow} backend and optimized with \textsc{Adam} using a batch size of 1000 and 50 epochs.
Training is performed over the QCD sample to obtain the calibration function.
The learned calibration function is then applied to both the QCD and BSM test samples.

The result of MSE calibration is shown in \Fig{MSE_fit}.
Prior to any calibration, the detector response is about 5\% low in both the QCD and BSM test samples.
After calibration, the mean is nearly unity for the QCD sample, albeit with a large width -- that is to say, the average bias is close to zero over the prior, but the average resolution is large.
For the BSM sample, though, the calibrated mean is far from unity, demonstrating the bias and prior dependence of the MSE calibration.
The MSE-based calibration obtained from the QCD fit is not universal, and gives poor results when applied to the BSM sample.%
\footnote{The converse is also true -- attempting to use a calibration fitted on the BSM sample will lead to bias on the QCD sample, or any other BSM sample for that matter. These non-universal fits lead to \emph{mass sculpting}, in which a fit depends strongly on the mass point used in training. See e.g.~\cite{Kitouni_2021} for discussions on sculpting and mass decorrelation.}

For comparison, in \Fig{GA_fit} we show results from a maximum-likelihood-based calibration trained on the QCD sample, using the Gaussian Ansatz in \Eq{gaussian_ansatz}.
The $A$, $B$, $C$, and $D$ networks of the Gaussian Ansatz each consist of three hidden layers with 32 nodes per layer, with the same activation functions, batch size, and epochs as in the Gaussian example.
The calibration function trained on the QCD sample can be used for the BSM sample, and as \Fig{GA_fit} shows, the calibration is indeed universal and unbiased, as expected.

\subsection{Data-based Calibration}

\begin{figure*}[t]
    \centering
    \subfloat[]{
        \includegraphics[width=0.42\textwidth]{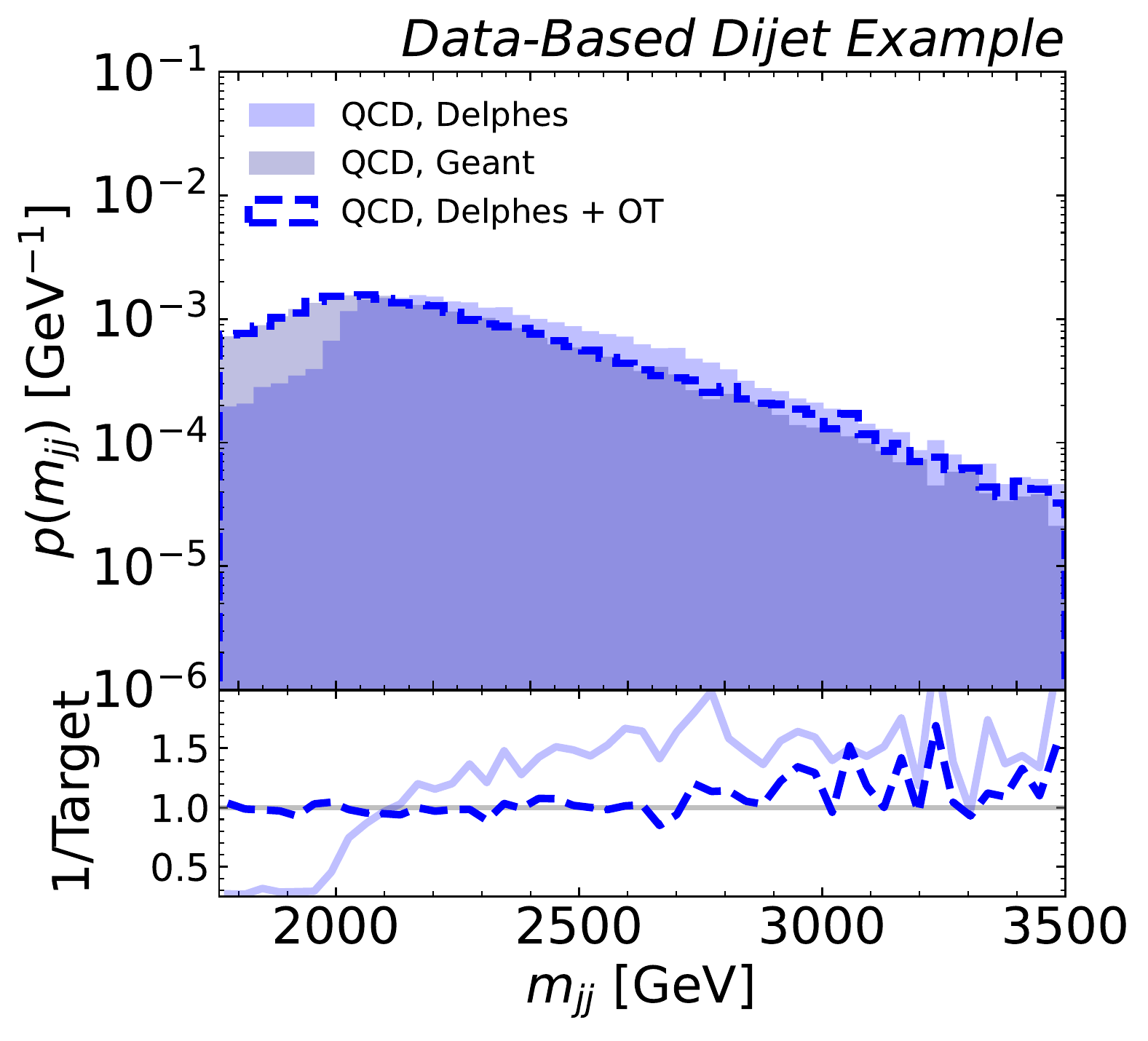}
        \label{fig:QCD_OT}
    }
    $\qquad$
    \subfloat[]{
        \includegraphics[width=0.42\textwidth]{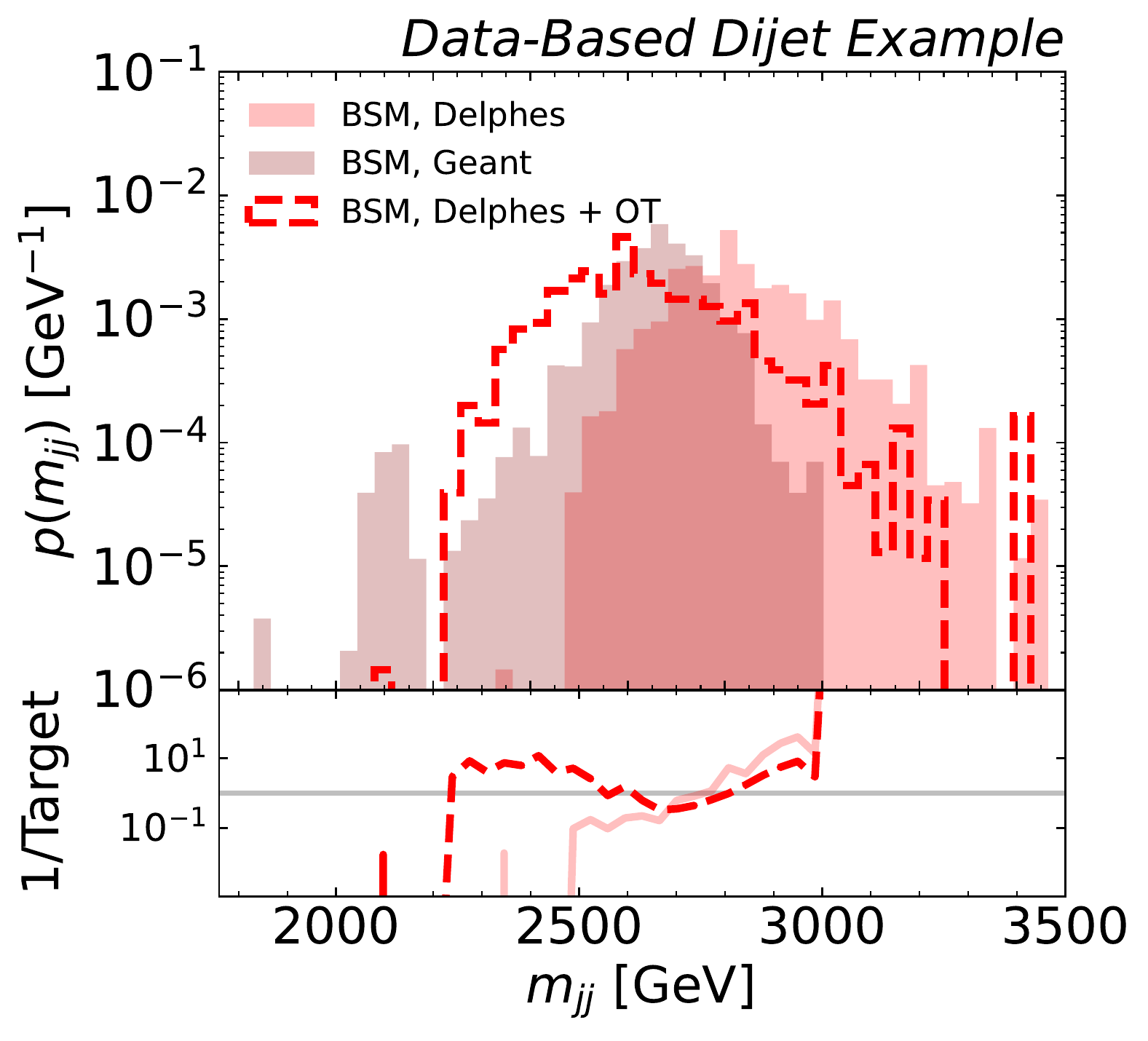}
        \label{fig:BSM_OT}
    }
    \caption{
    The reconstructed $m_{jj}$ for (a) QCD and (b) BSM events in the fast and full simulation, with and without the data-based OT calibration.
    The calibration is performed on the QCD sample, which closes, and the same calibration is applied to the BSM sample.
    Note that for the BSM sample, the ratio plot is in log-scale, indicating a very large bias.
    }
    \label{fig:dijet_OT}    
\end{figure*}

The goal for the data-based calibration task is to ``correct'' $p_{\rm sim}(m^{\text{reco}}_{jj})$, given by the fast simulation (\textsc{Delphes}), to the observed data distribution $p_{\rm data}(m^{\text{reco}}_{jj})$, given by the full simulation (\textsc{Geant4}).
We now apply the same procedure described in \Sec{dataGauss} to the dijet example.

An OT-based calibration is derived using QCD jets, to align the fast simulation \textsc{Delphes}) sample with the full simulation \textsc{Geant4} sample.
The calibration function, given by the optimal transport map (\Eq{OT_map_1D}), can be computed numerically by sorting and integrating the weighted data points to build the cumulative distribution functions.
On the QCD sample, this calibration closes by construction.
In particular, as shown in \Fig{QCD_OT}, the blue dashed line in the ratio plot fluctuates around unity, with deviations due to statistical fluctuations that differ between the two halves of the event samples.

When this calibration is applied to the BSM events, however, the calibration overshoots, as shown with the red dashed line in the ratio plot in \Fig{BSM_OT}.
While the resulting dashed distribution agrees better with the data histogram in dark red than does the fast sim histogram in light red, the overall agreement is still rather poor.
This again highlights the issue of prior dependence in data-based calibrations.

\section{Conclusions}
\label{sec:conclusions}

In this paper, we explored the prior dependence of machine-learning-based calibration techniques.
There is a growing number of machine learning proposals for simulation-based and data-based calibration and in nearly all cases, there is a prior dependence.
We highlighted the resulting calibration bias in a synthetic Gaussian example and a more realistic particle physics example of dijet production at the LHC.

In the simulation-based calibration case, most proposals learn a truth target from detector-level observables using loss functions like the MSE.
A neural network trained in this way will learn the average true value given the detector-level inputs, which depends on the spectrum of truth values.
However, we have shown that this will yield a calibration that lacks the critical properties of universality and closure.

There are fewer proposals for machine learning data-based calibrations, but we studied one recent idea based on OT and showed its prior dependence.
While we  focused on one-dimensional examples, the prior dependence is a generic feature of these approaches.
Going to higher dimensions may even exacerbate the issue since it is harder to visualize and control prior differences in many dimensions.

New learning approaches are required to ensure that machine learning-based calibrations are universal.
For simulation-based calibration, the ATLAS collaboration has proposed a prior-independent method called \textit{generalized numerical inversion}~\cite{ATL-PHYS-PUB-2018-013,ATL-PHYS-PUB-2020-001}.
While prior independent, this technique is typically biased and does not scale well to many dimensions.
We proposed a new approach based on maximum likelihood estimation in \Ref{frequentstway}, based on parametrizing the log-likelihood with a Gaussian Ansatz. 
Maximum-likelihood-based approaches are prior independent by construction and are well-motivated statistically.
Parametrizing the maximum likelihood estimator with neural networks requires a different learning paradigm than current approaches, but it extends well to many dimensions.
To our knowledge, there are currently no prior-independent data-based calibration approaches.

To make the most use of the complex data from the LHC and other HEP experiments, it is essential to use all of the available information for object calibration.
This will require modern machine learning to account for all of the subtle correlations in high dimensions.
It is important, however, that we construct these machine learning calibration functions in a way that integrates all of the features of classical calibration methods.
We highlighted prior independence in this paper as a cornerstone of calibration.
In the future, innovations that incorporate knowledge of the detector response or physics symmetries may further enhance the precision and accuracy of machine learning calibrations.

\section*{Code and Data}

The code for this paper can be found at \url{https://github.com/hep-lbdl/calibrationpriors
}, which makes use of \textsc{Jupyter} notebooks~\cite{Kluyver:2016aa} employing \textsc{NumPy}~\cite{harris2020array} for data manipulation and \textsc{Matplotlib}~\cite{Hunter:2007} to produce figures. All of the machine learning was performed on a Nvidia RTX6000 Graphical Processing Unit (GPU).  The physics datasets are hosted on \textsc{Zenodo} at~\Refs{komiske_patrick_2019_3341502,komiske_patrick_2019_3341770,komiske_patrick_2019_3341772,nachman_benjamin_2021_5108967}. 

\vspace{5mm}

\begin{acknowledgments}

BN is supported by the U.S. Department of Energy (DOE), Office of Science under contract DE-AC02-05CH11231.
RG and JT are supported by the National Science Foundation under Cooperative Agreement PHY-2019786 (The NSF AI Institute for Artificial Intelligence and Fundamental Interactions, \url{http://iaifi.org/}), and by the U.S. DOE Office of High Energy Physics under grant number DE-SC0012567.

\end{acknowledgments}

\bibliography{myrefs,HEPML}

\end{document}